\documentclass[12pt]{article}
\usepackage[mathscr]{eucal}
\usepackage{graphicx,subfigure}
\usepackage{bm}
\usepackage{bbold}
\usepackage{caption}
\usepackage{multirow}
\usepackage{amsmath}
\usepackage{textgreek}
\usepackage[greek,english]{babel}
\usepackage{xcolor}
\usepackage{graphicx}
\usepackage{empheq}
\usepackage{axodraw2}
\usepackage{amsfonts}
\usepackage{amssymb}
\usepackage{dsfont}
\usepackage{ytableau}
\usepackage{overpic}
\usepackage{hyperref} 
\usepackage{makeidx}
\usepackage{empheq}

\newcommand{\be}{\begin{eqnarray}}
\newcommand{\ee}{\end{eqnarray}}
\newcommand{\bc}{\begin{center}}
\newcommand{\ec}{\end{center}}
\newcommand{\nn}{\nonumber \\}
\newcommand{\lb}{\label}
\newcommand{\p}[1]{(\ref{#1})}

\renewcommand{\u}{\underline}

\begin{document}

\begin{titlepage}

\vspace*{0.2cm}

\begin{center}

{\LARGE\bf  QCD sum rules: Borel parameter vs. Euclidean time}

\vspace{2cm}

{\Large Andrei Smilga} \\

\vspace{0.5cm}

{\it SUBATECH, Universit\'e de
Nantes,  4 rue Alfred Kastler, BP 20722, Nantes  44307, France. }

\end{center}
\vspace{0.2cm} \vskip 0.6truecm \nopagebreak

   \begin{abstract}
We explore a modification of QCD sum rules where, instead of Borel transforms of current correlators, one considers the correlators in coordinate space as functions of Euclidean time.
Taking the nucleon channel as an example, we derive such Euclidean time sum rules and compare them with the traditional Borel sum rules. We show that a rough estimate of nucleon mass and residue is also possible working in coordinate space, but such sum rules are much more affected by the uncertainties in power corrections and continuum contribution than the  Borel ones: the fiducial interval is practically absent.

The Gaussian sum rules with the weight $\sim \exp\{-(s - \hat s)^2/\mu^4\}$ have roughly the same status as the Borel ones.
\end{abstract}

\end{titlepage}

\tableofcontents

\setcounter{footnote}{0}

\setcounter{equation}0
\newcommand{\vecind}[1]{\mbox{\scriptsize \boldmath $#1$}}
\newcommand{\ga}{\lower.7ex\hbox{$
\;\stackrel{\textstyle>}{\sim}\;$}}
\newcommand{\la}{\lower.7ex\hbox{$
\;\stackrel{\textstyle<}{\sim}\;$}}
\newcommand{\beq}{\begin{equation}}
\newcommand{\eeq}{\end{equation}}
\newcommand{\pd}{\partial}
\newcommand{\bebox}{\begin{empheq}}
\newcommand{\ebox}{\end{empheq}}
\renewcommand{\u}{\underline}
\renewcommand{\vec}[1]{{\bf #1}}
\newcommand{\red}{\color{red}}
\newcommand{\blue}{\color{blue}}

\section{Introduction} 
QCD sum rules is a powerful method to determine the properties of the lightest hadron states based on the first QCD principles. It capitalizes on the idea of quark-hadron duality. A correlator of some hadron currents can on one hand be evaluated by calculating certain QCD diagrams and on the other hand the same correlator can be represented as a sum over hadron states with the corresponding quantum numbers of which the lightest 
state provides the dominant contribution.

This method was mostly developped in the Institute for Theoretical and Experimental Physics (ITEP) in Moscow. In Ref. \cite{6auth}, it was successfully applied to describe the spectrum of charmonium. Then Shifman, Vainshtein and Zakharov elaborated this method including in the theoretical left-hand side of the sum rules nonperturbative contributions due to quark and gluon condensates \cite{SVZ1}. Doing this, they were able to evaluate the masses and residues of the light meson resonances ($\rho, \omega, A_1$) \cite{SVZ2}. Ioffe generalized this method for the nucleons and other baryons \cite{Ioffe-nucl} deriving in particular the famous {\it Ioffe formula}
 \be
m_N \ \approx \ (4\pi^2 |\langle \bar q q \rangle_{\rm vac}|)^{1/3}.
 \ee
Then this method was generalized to 3-point correlators. This allowed one to evaluate the electromagnetic formfactor of pion \cite{form} and other mesons \cite{formmes}, to calculate the magnetic moments of nucleons and other baryons \cite{magn-mom}, to determine the axial nucleon coupling constant $g_A$ \cite{gA} and derive many other important theoretical results related to the low-energy hadron physics. We are addressing the reader to the reprint volume \cite{sumrul} including the basic papers devoted to this subject.
 
\section{Correlator of nucleon currents in QCD}
To make the paper self-contained, we have to remind some well-known to  experts facts.
We follow Lecture 17 of Ref. \cite{Lectures}.

Following \cite{Ioffe-nucl}, consider the current
\be
\label{curIof}
\eta(x) \ =\ \epsilon^{ABC} \left[ u^A(x) {\cal C} \gamma_\mu u^B(x) \right]\gamma_5 \gamma^\mu d^C(x),
 \ee
 where $u(x)$ and $d(x)$ are the up and down quark fields, $A,\,B,\,C$ are the fundamental 
color indices, and  ${\cal C}$ is the charge conjugation matrix with the properties:\footnote{It follows:
$$ 
\bar\eta(x) \ =\ \epsilon^{ABC} \left[ \bar u^A(x)  \gamma_\mu {\cal C}\,\bar u^B(x) \right] \bar d^C(x)\gamma_5 \gamma^\mu.
$$}
$$
{\cal C}^2 =  {\cal C} {\cal C}^\dagger = \mathbb{1}, \qquad \gamma_0 ({\cal C} \gamma_\mu)^\dagger \gamma_0 = -\gamma_\mu {\cal C}, \qquad  {\cal C}\gamma_\mu {\cal C} = -\gamma_\mu^T.
$$
The spinorial 
fermionic current (\ref{curIof}) has a nonzero residue, 
 \be
\label{resnuc}
\langle 0|\eta|P\rangle \ =\ \lambda_N u_P,  
\ee
in the proton state.\footnote{Our conventions: ``P" is proton,
 ``p" is momentum, ``N" is nucleon, and ``n" is neutron.} 

\begin{figure}
\bc
\includegraphics[width=.6\textwidth]{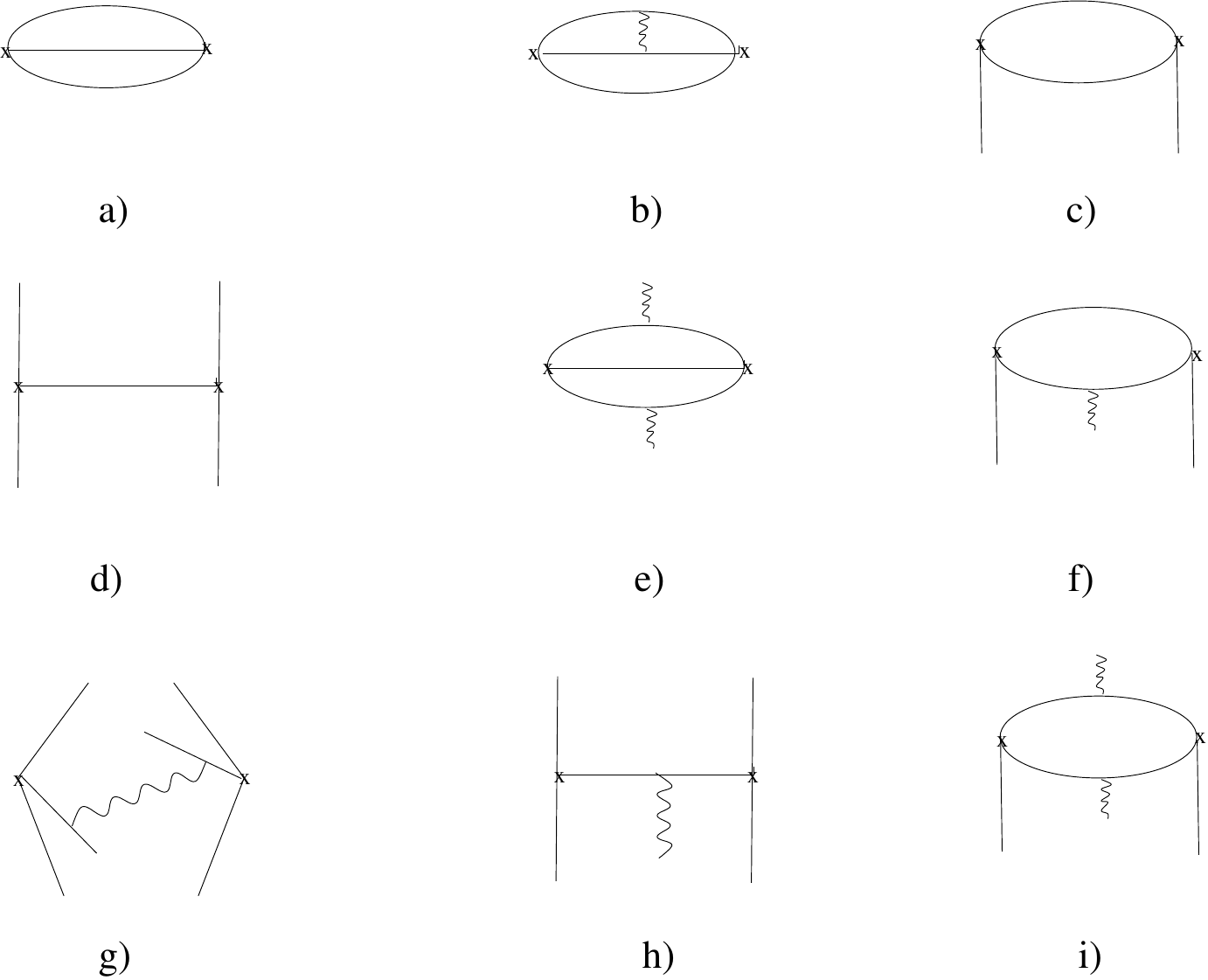}
\ec
        
\caption{Some QCD graphs contributing to $\Pi_{ij}(\tau)$.}
\label{fignucl}
\end{figure}

Consider now the Euclidean correlator $\Pi_{ij} (\tau) =\langle \eta_i(\tau) \bar \eta_j(0) \rangle$. 
The leading contribution at small $\tau$ is given by the graph in Fig.~\ref{fignucl}a. A simple calculation  gives

\be
 \label{Pipert}
 \left[ \Pi_{ij} (\tau) \right]^{\rm pert} \ =\ \frac{24  (\gamma_0)_{ij}}{\pi^6 \tau^9} 
\left\{1 + O[\alpha_s(\tau)] \right\} .
 \ee

The coefficient of $\alpha_s(\tau)$ turns out to be comparatively small and, in the region of interest  for the parameter $\tau$, the perturbative corrections to the leading-order result can be  disregarded.

There are, however, essential {\it nonperturbative} corrections.
 The most important ones are 
associated with the nonzero quark condensate
 $\Sigma = \langle u \bar u \rangle_0 =  \langle d \bar d \rangle_0 $. Technically, the presence of the latter
 can be taken into account by  writing the quark Green's functions in the form
 \be
 \label{Grqucond}
 \langle q_i^A(\tau) \bar q_j^B(0) \rangle \ =\ \delta^{AB} \left[
  \frac {(\gamma_0)_{ij}}{2\pi^2 \tau^3} + \frac \Sigma{12} \delta_{ij} \right]\ + \ \cdots
 , \ee
 where the dots stand not only for the perturbative corrections $\propto \alpha_s^n(\tau)$,
 but also for the higher nonperturbative corrections like 
$\propto \langle \bar q \sigma_{\alpha\beta} \hat G_{\alpha\beta} q \rangle_0$, $G$ being the gluon field.

It is important that  the higher nonperturbative  corrections involve extra factors $ \propto \tau^k$ with positive $k$. 
     When $\tau$ is small, they are all suppressed. Substituting the propagator 
(\ref{Grqucond}) 
into the loop in Fig.~\ref{fignucl}a, we obtain
\be
\label{PiSigma}
\Pi_{ij} (\tau) \ =\ \frac{24  (\gamma_0)_{ij}}{\pi^6 \tau^9} \left[1 + 
\frac{\pi^4 \Sigma^2 \tau^6}{72} \right] \ + \frac {2\Sigma}{\pi^4 \tau^6} \delta_{ij}
  . \ee
 The contributions $\propto \Sigma$ and $\propto \Sigma^2$ can be described by the graphs in 
Fig.~\ref{fignucl}c,d with disconnected quark lines.

The estimate (\ref{PiSigma}) takes into account the most essential 
{\it power corrections}
\index{power corrections} (they are suppressed at small $\tau$ as a power of $\tau \mu_{\rm hadr}$) to the leading order result
(\ref{Pipert}) and is already reasonable enough. It can be improved, however. 

 {\it (i)} As was mentioned above, perturbative corrections coming from the graphs like in 
Fig.~\ref{fignucl}b with the standard gluon propagator 
are not so important. But the gluon propagator involves, by the same token as the quark one, 
also
nonperturbative contributions. The leading contribution is  associated with the 
{\it gluon condensate}.
\index{condensate!gluon}  Choosing  the
Fock-Schwinger gauge \cite{Fock-Schw}, 
$x_\alpha A_\alpha^a (x) = 0$, we can express $A_\alpha^a(x)  = -
(1/2) x_\beta G^a_{\alpha\beta}+\cdots $  and represent the Euclidean gluon Green's functions as
 \be
\label{gluprcond}
 \langle A_\alpha^a(x) A_\beta^b(y) \rangle = \frac {i\delta^{ab} \delta_{\alpha\beta}}
{4\pi^2 (x-y)^2} 
+ \frac {\delta^{ab} \langle G^c_{\mu\nu} G^c_{\mu\nu}\rangle_0 }{48(N_c^2 -1)} 
[\delta_{\alpha\beta}
(xy) - y_\alpha x_\beta ] + \cdots \nonumber\\
 \hspace{-7cm}\ee

The contribution of the second term in Eq.~(\ref{gluprcond}) to the
nucleon polarization operator is usually represented by the graphs
with a disconnected gluon line like
in Fig.~\ref{fignucl}e. The gluon condensate
 $\langle G^2 \rangle_0$
is a universal parameter characterizing the vacuum properties. It  enters not only the
nucleon sum rules discussed here, where it gives a comparatively small correction, but also
the ITEP sum rules for all other hadron channels. The value of the gluon
\index{condensate!gluon} condensate is best
determined from the sum rules in the charmonium channel \cite{SVZ2}:
\be
\label{glucond}
\left\langle \frac {\alpha_s}{\pi}   G^c_{\mu\nu} G^c_{\mu\nu} \right\rangle_{0} \ \approx 
\ 0.012 \ {\rm GeV}^4
 \ee
with an uncertainty $\approx$ 30--40 \%.

{\it (ii)} Another correction is brought about by the nonzero vacuum expectation value 
  \be
\label{qsigGq}
\langle g_s i\bar q \sigma_{\alpha\beta} \hat G_{\alpha\beta}  q \rangle_{0}   \ \stackrel{\rm def}{=} m_0^2\langle \bar  q  q \rangle_{0}   = - m_0^2 \Sigma,
\ee
with  $\sigma_{\alpha\beta} = \gamma_{[\alpha} \gamma_{\beta]}$. The corresponding contributions to the correlator  $\Pi_{ij}(\tau)$ are due to the graph in Fig. \ref{fignucl}f and also Fig. \ref{fignucl}c, where a nonperturbative contribution due to the average \p{qsigGq} is taken into account. The two contributions $ \propto m_0^2 \Sigma$ to the nucleon correlator happen to cancel, however. 

{\it (iii)} Besides, there is a nonvanishing contribution of the average $ m_0^2 \Sigma^2$ described by the graphs  in Fig. \ref{fignucl}d and Fig. \ref{fignucl}h, the contribution $\propto \langle G^2 \rangle_0 \Sigma$ described by the graph in Fig.   \ref{fignucl}i and the contribution $\propto \alpha_s \Sigma^3$ due to the graph in Fig. \ref{fignucl}g.

Putting everything together, we obtain:
 \be
  \label{LHSnuc}
  \!\!\!\!\!\!\!\!\! \Pi_{ij} (\tau) \ &=&\ \frac{24 (\gamma_0)_{ij}}{\pi^6 \tau^9} \left[1 + 
  \frac {b\tau^4}{3\cdot 2^8} +
\frac{\pi^4 \Sigma^2 \tau^6}{72}   - \frac{\pi^4 m_0^2 \Sigma^2 \tau^8}{9 \cdot 64} \right]\hspace{.8cm}   
\nonumber \\
&+& \ \frac {2\Sigma}{\pi^4 \tau^6} \delta_{ij} \left[1 - 
  \frac {b\tau^4}{3\cdot 2^8} \right]  \ 
-\  \frac{68}{81}  \delta_{ij}  \left( \frac{\alpha_s}{\pi} \right) \Sigma^3 \ln (\tau \mu_{\rm hadr}),
\ee
where $b = 4\pi^2 \langle (\alpha_s/\pi) G^2 \rangle \approx .47 \,{\rm GeV}^4$.

\section{Borel sum rules}

An attempt to determine the nucleon properties using \p{LHSnuc} will be performed a bit later, and in this section, we will explain how it is done in a standard way from Borel sum rules. We follow Ref. \cite{BelIof}. 

Consider the Fourier image  of the
coordinate space polarization operator (\ref{LHSnuc}),  go over in the Minkowski space and
take the  imaginary part of $\Pi(p)$ 
in the timelike region $s = p^2 \geq 0$. We obtain
\be
\lb{ImPi}
\frac 1\pi {\rm Im}\, \Pi(s)\ =\ \rho_1(s)  \hat p + \rho_2(s)
\ee 
with 
\be
\lb{ro}
\rho_1(s) \ =\  \frac {2s^2 + b}{128\pi^4} +
 \frac{2\Sigma^2}3 \delta(s) + \frac {m_0^2 \Sigma^2}6 \delta'(s) , \nn
\rho_2(s) \ =\ \frac \Sigma{4\pi^2} \left[ s - \frac b{18} \delta(s) \right] -
 16 \pi^2 \frac{17}{81} \left( \frac {\alpha_s}\pi \right) \Sigma^3 
\delta'(s).
\ee
Calculate now the {\it Borel} (alias, Laplace)  transform of the polarization operator:
\be
\label{Borel}
 {\cal B}_{M^2} \Pi(p) \ =\ \frac 1 \pi \int_0^\infty  ds \ e^{-s/M^2}
[/\!\!\!{p} \rho_1(s) + \rho_2(s) ] . 
\ee
This gives the left-hand side of the sum rule. For the right-hand side, we have to express the spectral densities $\rho_{1,2}(s)$ as a sum over the hadron states with the nucleon quantum numbers. To begin with, it is the nucleon itself:
$$\rho_1^N(s) = \lambda^2 \delta(s- m^2), \qquad \rho_2^N(s) = m \lambda^2 \delta(s- m^2), 
$$
where $\lambda$ is the residue defined in \p{resnuc} and $m$ is the nucleon mass. Besides, there are many excited states. They give a sum over the broad Breit-Wigner peakes in the spectral density. For large enough $s$, all such peaks coalesce into a smooth curve. Due to asymptotic freedom, this is nothing but the quark spectral density, the imaginary part of the diagram in Fig. \ref{fignucl}a with a correction from the diagram Fig. \ref{fignucl}e.
We may adopt the {\it resonance + continuum model}:
\be
 \label{rescont}
 \rho_1(s) \ =\ \lambda^2 \delta(s - m^2) + \rho_1^{\rm quark}(s)
 \theta(s - W_1^2) \ ,\nonumber \\
 \rho_2(s) \ =\ m\lambda^2 \delta(s - m^2) + \rho_2^{\rm quark}(s)
 \theta(s - W_2^2),
  \ee
  where $W_{1,2}$ are certain thresholds,  free parameters of the model.
The simplest assumption is $W_1 = W_2$ and we will stick to it. In numerical estimates, we will take the value $W = 1.5 \,{\rm GeV}$

The Borel sum rule is derived when the model \p{rescont} is substituted in the dispersive integral \p{Borel}  and this is compared with the same integral for the theoretical spectral densities \p{ro}.  We obtain:\footnote{We took the result from Ref. \cite{BelIof}, but disregarded for simplicity the effects due to anomalous dimensions of the operators. Taking them into consideration increase a little bit the predictions for the nucleon residue, but otherwise  these effects do not play an important role.}
\be
\lb{BorelSR}
 M^6    + \frac 43 a^2  + \frac 14 bM^2 - 
 \frac { a^2 m_0^2}{3M^2}  
  &=&   \tilde \lambda^2 \exp(-m^2/M^2)  
 +  M^6 E_2(W^2/M^2) + \frac {bM^2}4  E_0(W^2/M^2) 
 , \nn
2a M^4   \ +\ \frac {136}{81} \left( \frac {\alpha_s}\pi \right) 
\frac {a^3}{M^2} \ -\ \frac {ab}{9}  &=&   m \tilde \lambda^2 \exp 
(- m^2/M^2)  + 2a M^4  E_1(W^2/M^2)\ ,
 \ee 
  where\  \ $a = 4\pi^2 \Sigma \approx .55 \,{\rm GeV}^3$\,, \ \ $\tilde \lambda^2 = 32 \pi^4 \lambda^2$ and
 $$E_0(x) = e^{-x}, \quad 
  E_1(x)  =  (1 + x )e^{-x}, \quad   E_2(x)  = (1 + x + x^2/2)e^{-x}. $$ 
The Borel parameter $M^2$ plays   a regulator role, similar to the role of the hot water tap in the bathroom. When $M^2$ is small, the continuum contribution is suppressed and the sum rule is focused on the nucleon. On the other hand, the power corrections (including the corrections that we cannot evaluate) acquire an overwhelming importance in this case and the expressions \p{BorelSR} cannot be trusted --- the water is too cold.

And if $M^2$ is large, the power corrections are suppressed, only the diagram in   \ref{fignucl}a contributes to the spectral density, but the dominant contribution to the integral \p{Borel} of the phenomenological spectral density comes from the excited states (the water is too hot) and is not sensitive to the nucleon contribution. Thus, one should keep the water lukewarm and  choose $M^2$ not too large and not too small so that both nonperturbative power corrections and the excited state contribution [described by the model \p{rescont}] are under control.  The question is whether such {\it fiducial interval} of $M^2$ exists.

Assuming that it does (we will justify it soon), we can play, to begin with, the following game:
\begin{enumerate}
 \item Neglect the continuum contributions in Eqs.~(\ref{BorelSR}) 
and set $E_0 = E_1 = E_2  = 1$.

\item  Neglect the terms $\propto b,\ \propto m_0^2$, and $\propto \alpha_s a^3$. 
 \item Set $M = m$ and compare the two sum rules. We obtain the equation
\be
\label{eqform}
m^6 + \frac 43 a^2 \ =\ 2 a m^3.
 \ee
\end{enumerate}
The equation (\ref{eqform}) has no real solutions for $m$ (a real solution  appears, however,  
if taking into account the neglected terms), but the best agreement 
(a rather close one) is achieved at $m = a^{1/3}$, which is the Ioffe formula.

Take now the value of $M$ such that the leading power correction $\propto a^2$ in the LHS of the first sum rule in Eq. \p{BorelSR} is a half of the main perturbative contribution. In other words, $M^6 = 8a^2/3$. This gives $M \approx   .96$ GeV. If so, the relative continuum contribution in the sum rule is
 \be
\lb{frac-cont}
R_{\rm cont} \sim \frac {M^6 E_2(W^2/M^2)}{M^6 + M^6/2 + b M^2/4 -  ( a^2 m_0^2)/(3M^2) } \ \approx \ .36.
 \ee
Continuum contribution constitutes about one third of the whole contribution in the RHS,  The relative contribution of the power corrections in the LHS is
\be
R_{\rm power} \sim \frac{M^6/2 + bM^2/4 -   a^2 m_0^2/(3M^2)  }{M^6 + M^6/2 + bM^2/4 -   a^2 m_0^2/(3M^2)} \approx .38,
\ee
which is about the same. This is acceptable, both contributions are roughly under control. A fiducial interval exists.

\begin{figure}
\bc
\includegraphics[width=.8\textwidth]{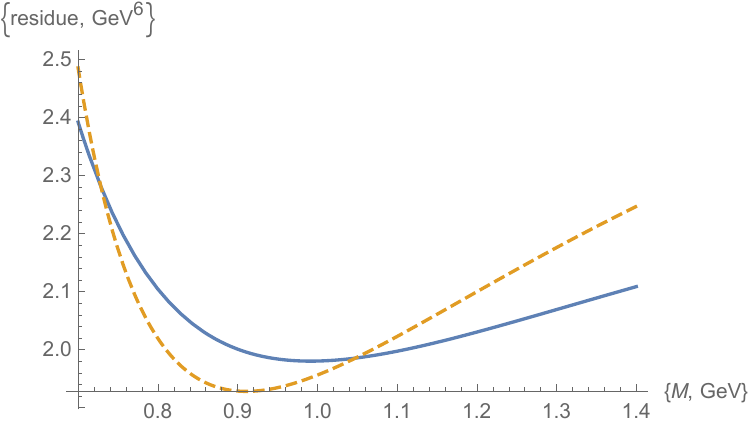}
\ec
        \caption{Nuclear residue as a function of Borel parameter, as follows from the first sum rule in Eq. \p{BorelSR} (solid line) and from the second one (dashed line).  $m = 940$ MeV.}
\label{plotres}
\end{figure}

To check the consistency of the two sum rules in \p{BorelSR}, we may take the experimental value of the nucleon mass, $m = .94$ GeV and plot the dependence of the residue $\tilde \lambda^2$ on $M^2$, as follows from the first sum rule and from the second. As we see from  Fig. \ref{plotres}, the consistency is there. The value of residue $\tilde \lambda^2 \approx 2 $ GeV$^6$ is somewhat underestimated, however, due to the fact that we neglected the anomalous dimension factor in our analysis. If taking it into consideration as was done is Ref. \cite{BelIof}, we would rather obtain $\tilde \lambda^2 \approx 3 $ GeV$^6$.
 
It is instructive to repeat this exercise with other assumed values for the nucleon mass. In Figs. \ref{Bor08}, \ref{Bor11}, we presented the results for $m = 800$ MeV and for $m = 1100$ MeV. As one can see,  the agreement between the chirality-odd and chirality-even sum rules 
   is considerably worse in both cases. We conclude that the Borel sum rules allow one to evaluate the nucleon mass with the accuracy 10--15 \%.

 \begin{figure}[ht!]
 \lb{plotres-other}
     \begin{center}

        \subfigure[$m = 800$ MeV]{
           \lb{Bor08}  \includegraphics[width=0.45\textwidth]{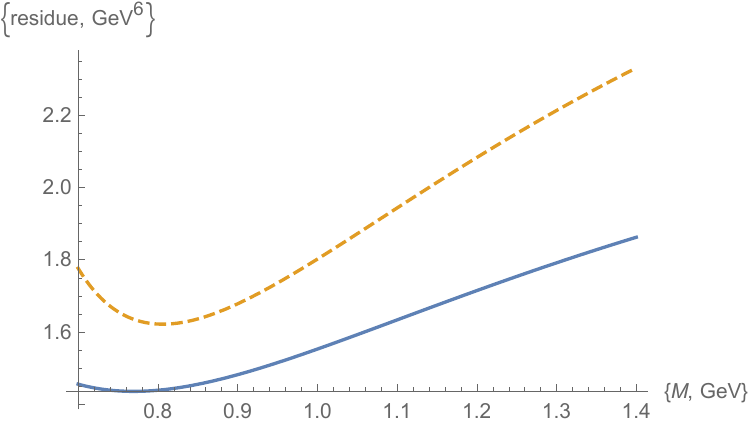}
        }
        \subfigure[$m = 1100$ MeV]{
         \lb{Bor11}
           \includegraphics[width=0.45\textwidth]{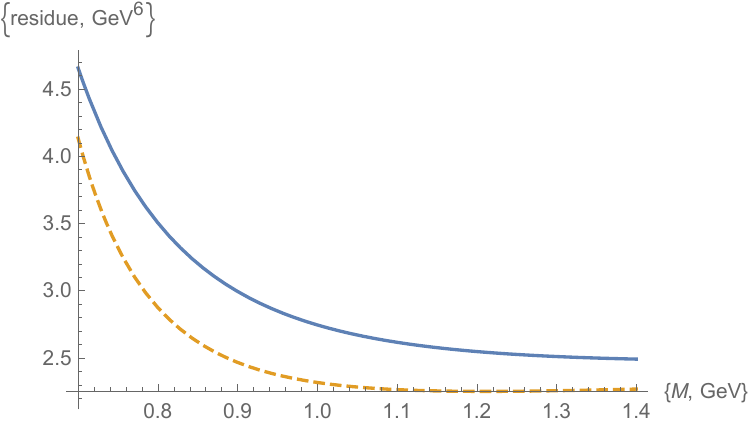}
        }
    \end{center}
    \caption{The same as in Fig. \ref{plotres} with other trial nucleon masses. }
    \end{figure}

\section{Euclidean time sum rules}

The Euclidean Green's function of the nucleon, a massive spinor particle, has the form

\be
\label{Grfermass}
G_N(\tau) \ =\ \frac {m^2}{4\pi^2 \tau} [  \gamma_0 K_2(m\tau) + 
 K_1 (m\tau)],
 \ee
 $K_n(x)$ being the Macdonald  functions. And the Euclidean correlator $\Pi(\tau)$ is expressed via the spectral densities  in  Eq. \p{ImPi} as  
 \be  
   \label{spec12}
  \Pi^{\rm phys}(\tau) & =& \int_0^\infty  ds\ \rho_1(s) 
\left[ \frac{\gamma_0 s}{4\pi^2\tau} K_2
 (\sqrt{s} \tau) \right] \nonumber \\
 & + & \int_0^\infty ds\ \rho_2(s) \left[ \frac{\sqrt{s}}{4\pi^2\tau} 
K_1 (\sqrt{s} \tau) \right].
\ee
Again, we compare these dispersive integrals for the QCD spectral densities \p{ro} with the same integrals for the phenomenological spectral densities \p{rescont}. We obtain two sum rules:\footnote{Besides Ref. \cite{Lectures}, similar formulas for meson correlators were written in Ref. \cite{Zakharov}, but these authors did not try to perform a numerical fit.}

\be
  \label{sumrulnuc}
 &&\frac{96}{\pi^4 \tau^8}  + \frac b{8\pi^4 \tau^4} + \frac{4\Sigma^2}{3\tau^2} - 
\frac{m_0^2 \Sigma^2}6 \ =\ \lambda^2 m^2 
K_2(m\tau)  +  \frac{96}{\pi^4 \tau^8} Q_1(W)\ ,\ \ \ \ \ \ \ \ \nonumber \\ 
&&\frac{8\Sigma}{\pi^2 \tau^5}  - \frac {\Sigma b}{96\pi^2 \tau}
-  \frac{272}{81} \pi^2 \left( \frac{\alpha_s}{\pi} \right) \tau \Sigma^3 \ln (\tau \mu_{\rm hadr}) 
= \lambda^2 m^2 
K_1(m\tau) + \frac{8\Sigma}{\pi^2 \tau^5} Q_2(W), \ \ \ \ \ \ 
\ee
where
\be
\label{Q12}
Q_1(W) &=& \frac {\tau^8}{3\cdot 2^{11}} \int_{W^2}^\infty s^3 ds K_2(\sqrt{s} \tau) 
= \frac {(\tau W)^7}{3\cdot 2^{10}} K_3(\tau W) 
+ \frac {(\tau W)^6}{3\cdot 2^{8}} K_4(\tau W) + \frac {(\tau W)^5}{3\cdot 2^{7}} K_5(\tau W)   \ ,\nonumber \\ 
Q_2(W) &=&\frac {\tau^5}{32} \int_{W^2}^\infty s^{3/2} ds K_1(\sqrt{s} \tau)   
=\frac {(\tau W)^4}{16} K_2(\tau W)  + \frac {(\tau W)^3}{8} K_3(\tau W)  
 . \ee
The parameter $\tau$   plays in these sum rules exactly the same role as the parameter $M^{-1}$
in the Borel sum rules \p{BorelSR}:
 when $\tau$ is small, 
the power corrections
 on the theoretical side of the sum rules are suppressed while, 
 when $\tau$ is large, the
contribution of the excited states in the phenomenological side is suppressed. 

\begin{figure}
     \begin{center}
 \includegraphics[width=0.8\textwidth]{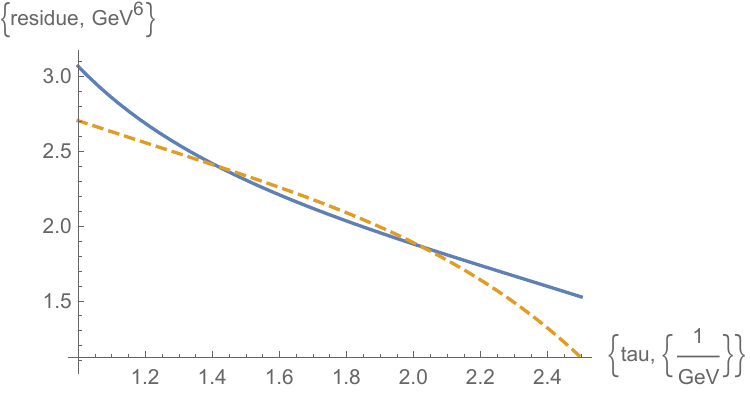}
    \end{center}
 \caption{Residues extracted from the Euclidean time sum rules \p{sumrulnuc} with $m = 800$ MeV.}
\label{hren}
    \end{figure}

The question now is whether a fiducial interval of $\tau$ where both power corrections and continuum contribution stay under control exists for Euclidean time sum rules, as it does for the Borel sum rules. And the  answer to this question seems to be negative. Let us try to proceed in the same way as we did for the Borel sum rules and find the value of $\tau$ where the power correction $\propto \Sigma^2$ constitute a half of the main perturbative term. It is equal to
\be 
\tau^* \ = \ \sqrt[3]{\frac 6{\pi^2 \Sigma}} \approx 3.5 \ {\rm GeV}^{-1}.
 \ee
This value seems to be large, but it is not large enough to focus the sum rule on the nucleon. The relative continuum contribution defined in the same way as in Eq. \p{frac-cont} is equal in the case to $\approx .75$, three times greater than the nucleon contribution.

Besides, at so large $\tau$, the higher power correction $\sim m_0^2 \Sigma^2$ becomes as large and even larger than the correction $\sim \Sigma^2$. It is negative and, if we increase $\tau$ a little more, it becomes dominant and soon the whole left-hand side of the sum rule flips sign.

The main reason for such a setback is the fact that the Bessel weights $K_{1,2}(\sqrt{s} \tau)$ behave as $\exp\{-\sqrt{s}\tau\}$ in the asymptotics, in contrast to the weight $\exp\{-s/M^2\}$ in the Borel integral \p{Borel}. As a result, the excited states are not suppressed enough. 

This notwithstanding, one can still extract some information on the nucleon mass and residue from the sum rules \p{sumrulnuc}. Let us do the same as we did for the Borel sum rules, fix the nucleon mass at some value, substitute it in \p{sumrulnuc}, solve the first and the second equation  for $\lambda_N$ and plot the results as functions of $\tau$. As is seen in Fig. \ref{hren}, these functions are practically equal and do not change much in the interval $1.5 \, {\rm GeV}^{-1} \la \tau \la 2 \, {\rm GeV}^{-1}$ if we fix the value of mass $m = 800$ MeV. The value of the residue thus extracted turns out to be close to the value $\tilde \lambda_N^2 \sim 2 \, {\rm GeV}^6$ as follows from the Borel sum rules. We want to emphasize, however, that the interval $1.5 \, {\rm GeV}^{-1} \la \tau \la 2 \, {\rm GeV}^2$ is not so much fiducial: the contribution of continuum rather than that of the nucleon pole dominates there.

 To conclude this section, one can observe that Euclidean time sum rules represent an attractive alternative to Borel sum rules suggested in Ref.\cite{SVZ1}: after all, Euclidean time $\tau$ is a more
natural and physical  quantity than the Borel parameter $M$, which looks somewhat artificial. But, being theoretically  transparent, these sum rules are not really suitable for practical applications because a fiducial interval of $\tau$ cannot be established. At least, it is so in the nucleon channel and one can guess that it is also so for other light hadrons.

\section{Gaussian sum rules}

In addition to the Borel sum rules and Euclidean time sum rules, one can also use other similar instruments to study hadronic spectrum. In particular, one can mention the {\it Gaussian} sum rules where the weight in the spectral integral is chosen not as 
$\exp\{-s/M^2\}$ in Eq. \p{Borel} and not as Macdonald functions in Eq. \p{spec12}, but in the form \cite{Gauss}
\be
\lb{weight}
f_{\mu, \hat s}(s) \ =\ \exp \left\{- \frac {(s - \hat s)^2}{\mu^4}   \right\}.
 \ee
Probably the most natural choice corresponds to $\hat s =0$. Such a weight suppresses the continuum better than the Laplace weight and, of course, better than the Euclidean time weight. The corresponding sum rules read:
\be
\lb{hyp-Borel}
\frac {\mu^6}4 \left[ {\rm erf}(\kappa) - \frac {2\kappa}{\sqrt{\pi}} e^{-\kappa^2}  \right]
+ \frac {b \mu^2}4 {\rm erf}(\kappa) + \frac {8 a^2}{3\sqrt{\pi}} \ &=&\ \frac 2{\sqrt{\pi}} \tilde \lambda^2 e^{-m^4/\mu^4} \nn
a\mu^4 (1 - e^{-\kappa^2}) - \frac {ab}9 \ &=&\ m  \tilde \lambda^2 e^{-m^4/\mu^4}
 \ee
with $\kappa = W^2/\mu^2$.

\begin{figure}
     \begin{center}
 \includegraphics[width=0.8\textwidth]{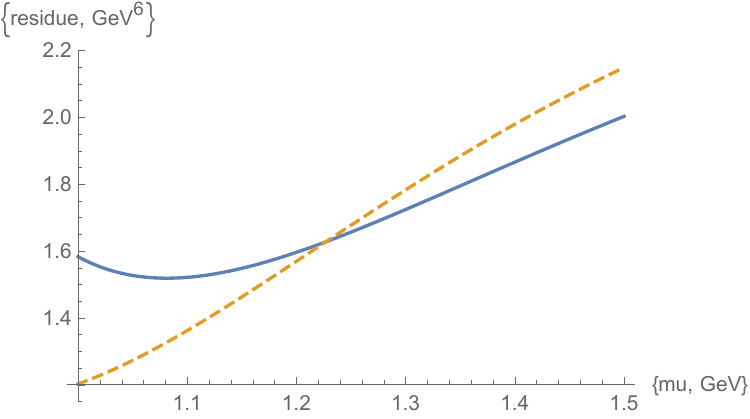}
    \end{center}
 \caption{Residues extractes from the Gaussian sum rules \p{hyp-Borel}  with $m = 940$ MeV.}
\label{plotres-hyp}
    \end{figure}

To find out whether a fiducial interval for the parameter $\mu^2$ exists, we proceed in the same way as we did for other sum rules.
 \begin{itemize}
\item  We consider the LHS for the first sum rule in \p{hyp-Borel} in the limit $\kappa \to \infty$ such that the continuum contribution is not subtracted:
$$  {\rm LHS}(\kappa = \infty) \ =\ \frac {\mu^6}4 + \frac {b\mu^2}4 + \frac {8a^2}{3\sqrt{\pi}}. $$
\item Then we choose the value of $\mu$ such that the power correction $\propto a^2$ is equal  $\mu^6/8$,  a half of the leading term. This gives $\mu \approx 1.24$ GeV corresponding to $\kappa \approx 1.46$. Then
$$ {\rm LHS}(\kappa) \ \approx \  .77 \, {\rm LHS}(\kappa = \infty),
$$
i.e. the continuum contribution is only 23 \% of the leading term in the LHS. This seems to be fine: a fiducial interval is there, as was the case for the Borel sum rules.

However, the plots for the nuclear residue as a function of $\mu^2$ that follows from 
\p{hyp-Borel} (see Fig. \ref{plotres-hyp}) are not so nice as the Borel plots in Fig. \ref{plotres}.     It seems to stem from the fact that the power correction coming from the 8-dimensional operator $(\bar q \sigma G q)(\bar q q)$ in the first sum rule and the correction
$\propto \alpha_s \langle \bar q q  \rangle^3$ in the second sum rule, which played a comparatively essential role for the Borel sum rules (the agreement with experiment and between themselves   would be a little worse without taking them into account), do not contribute in the sum rules with the weight $\exp\{-s^2/\mu^4\}$. Indeed, the corresponding contributions to the spectral densities in \p{ro} include the factor $\delta'(s)$ and 
$$ \int ds \, \delta'(s)\,  e^{-s^2/\mu^4} \ =\ 0. 
$$   

The results should   expected to be better if the weight \p{weight} with a nonzero $\hat s$ is chosen, so that the integral of $ \delta'(s)$ with this weight is nonzero. I am addressing the reader to Ref. \cite{Gauss-res} where such a study was conducted.
\end{itemize}

I wish to thank Makoto Oka for useful correspondence.


\begin{thebibliography}{96}

\bibitem{6auth} V.A. Novikov et al, {\it Charmonium and gluons: basic experimental facts and theoretical introduction}, Phys. Rept. {\bf 41} (1978) 1. 

\bibitem{SVZ1} M.A. Shifman, A.I. Vainshtein and V.I. Zakharov, {\it QCD and resonance physics. Theoretical foundations},
Nucl. Phys. {\bf B147}(1979) 385;

\bibitem{SVZ2} M.A. Shifman, A.I. Vainshtein and V.I. Zakharov, {\it QCD and resonance physics. Applications},
Nucl. Phys. {\bf B147}(1979) 448.

\bibitem{Ioffe-nucl} B.L. Ioffe, {\it Calculation of baryon masses in quantum chromodynamics}, Nucl. Phys. {\bf B188} (1981) 317,  {\bf B191} (1981) 591 (E).

\bibitem{form} B.L. Ioffe and A.V. Smilga, {\it Pion formfactor at intemediate momentum transfer in QCD},  Phys. Lett. {\bf 114B} (1982) 353;

 V.A. Nesterenko and A.V. Radyushkin, {\it Sum rules and the pion formfactor in QCD}, Phys. Lett. {\bf 115B} (1982).

\bibitem{formmes} B.L. Ioffe and A.V. Smilga,  {\it Meson widths and form factors at intermediate momentum transfer in non-perturbative QCD},
Nucl. Phys. {\bf B216} (1983) 373.

\bibitem{magn-mom} B.L. Ioffe and A.V. Smilga, {\it Nuclear magnetic moments and magnetic properties of the vacuum in QCD}, Nucl. Phys. {\bf B232} (1984) 109;

I.I. Balitsky and A.V. Yung, {\it Proton and neutron magnetic moments from QCD sum rules}, Phys. Lett. {\bf 129B} (1983) 328;

B.L. Ioffe and A.V. Smilga, {\it Hyperon magnetic moments in QCD}, Phys. Lett. {\bf 133B} (1983) 436.

\bibitem{gA} V.M. Belyaev and Ya.I. Kogan, {\it Quantum chromodynamics calculation of $g_A$}, JETP Lett. {\bf 37} (1983) 730.

\bibitem{sumrul}
 M. Shifman  (ed.)  {\it Vacuum structure and QCD sum rules}: reprint volume, North Holland, Amsterdam, 1992 .

\bibitem{Lectures} A. Smilga, {\it Lectures on Quantum Chromodynamics}, World Scientific, 2001. 

\bibitem{Fock-Schw}
V. Fock, {\it Proper time in classical and quantum mechanics}, Sov. Phys. {\bf 12} (1937) 404;

J. Schwinger, {\it On gauge invariance and vacuum polarization}, Phys. Rev. {\bf 82} (1951) 664. 

\bibitem{BelIof} V.M. Belyaev and B.L. Ioffe, {\it Determination of baryon and baryon-resonance masses from quantum chromodynamics
sum rules. Nonstrange baryons}, Sov. Phys. JETP {\bf 56} (1982) 493.

\bibitem{Zakharov} S. Narison and V.I. Zakharov, {\it Hints on the power corrections from current correlators in $x$ space }, Phys. Lett. {\bf B522} (2001) 266.

\bibitem{Gauss} R.A. Bertlmann, I. G. Launer  and E. de Rafael, {\it Gaussian sum rules in quantum chromodynamics and local duality}, Nucl. Phys. {\bf B250} (1985) 61.

\bibitem{Gauss-res} K. Ohtani, P. Gubler and M. Oka, {\it A Bayesian analysis of the nucleon QCD sum rules}, Eur. Phys. J. A. {\bf 47} (2011) 114;
{\it Parity projection of QCD sum rules for the nucleon}, Phys. Rev. {\bf D87} (2013) 034027.








\end{thebibliography}
\end{document}